\documentclass{article}
\usepackage{subfiles}
\usepackage{authblk}
\usepackage{graphicx}
\usepackage{amsmath,amssymb,amsfonts}
\usepackage{algorithmic}
\usepackage{textcomp}
\usepackage{xcolor}
\usepackage{soul}
\usepackage{url}
\usepackage{siunitx}
\usepackage{authblk}
\usepackage[labelfont=bf]{caption}
\usepackage[capitalise]{cleveref}
\usepackage[labelformat=simple]{subcaption}

\usepackage[sorting = none, backend = biber, style=numeric-comp]{biblatex}
\usepackage{graphicx}
\usepackage{soul}
\usepackage{xcolor}
\usepackage{amssymb}
\usepackage{geometry}
\usepackage{subcaption}
\usepackage{svg}
\usepackage{multirow}
\title{Real-time all-optical signal equalisation with silicon photonic recurrent neural networks}
\author[1,2]{Ruben Van Assche}
\author[1,2]{Sarah Masaad}
\author[1,2]{Emmanuel Gooskens}
\author[1,2]{Stijn Sackesyn}
\author[2,3]{Joris Van Kerrebrouck}
\author[2,3]{Xin Yin}
\author[1,2]{Peter Bienstman}

\affil[1]{Photonics Research Group, INTEC, Ghent University - imec, 9052 Ghent, Belgium.}
\affil[2]{imec, Kapeldreef 75, 3001 Leuven, Belgium.}
\affil[3]{IDLab, INTEC, Ghent University–imec, 9052 Ghent, Belgium.}

\date{}
\begin{document}
\twocolumn[
\begin{@twocolumnfalse}
\maketitle

\begin{abstract}
Communication through optical fibres experiences limitations due to chromatic dispersion and nonlinear Kerr effects that degrade the signal. Mitigating these impairments is typically done using complex digital signal processing algorithms. However, these equalisation methods require significant power consumption and introduce high latencies. Photonic reservoir computing (a subfield of neural networks) offers an alternative solution, processing signals in the analog optical domain. In this work, we present to our knowledge the very first experimental demonstration of real-time online equalisation of fibre distortions using a silicon photonics chip that combines the recurrent reservoir and the programmable readout layer. We successfully equalize a 28 Gbps on-off keying signal across varying power levels and fibre lengths, even in the highly nonlinear regime. We obtain bit error rates orders of magnitude below previously reported optical equalisation methods, reaching as low as \num{4e-7}, far below the generic forward error correction limit of \num{5.8e-5} used in commercial Ethernet interfaces. Also, simulations show that simply by removing delay lines, the system becomes compatible with line rates of 896 Gpbs. Using wavelength multiplexing, this can result in a throughput in excess of 89.6 Tbps. Finally, incorporating non-volatile phase shifters, the power consumption can be less than 6 fJ/bit.
\\
\end{abstract}
\end{@twocolumnfalse}
]
\section{Introduction}

\begin{figure}
     \centering
     
         \centering
         \includegraphics[width=0.5\textwidth]{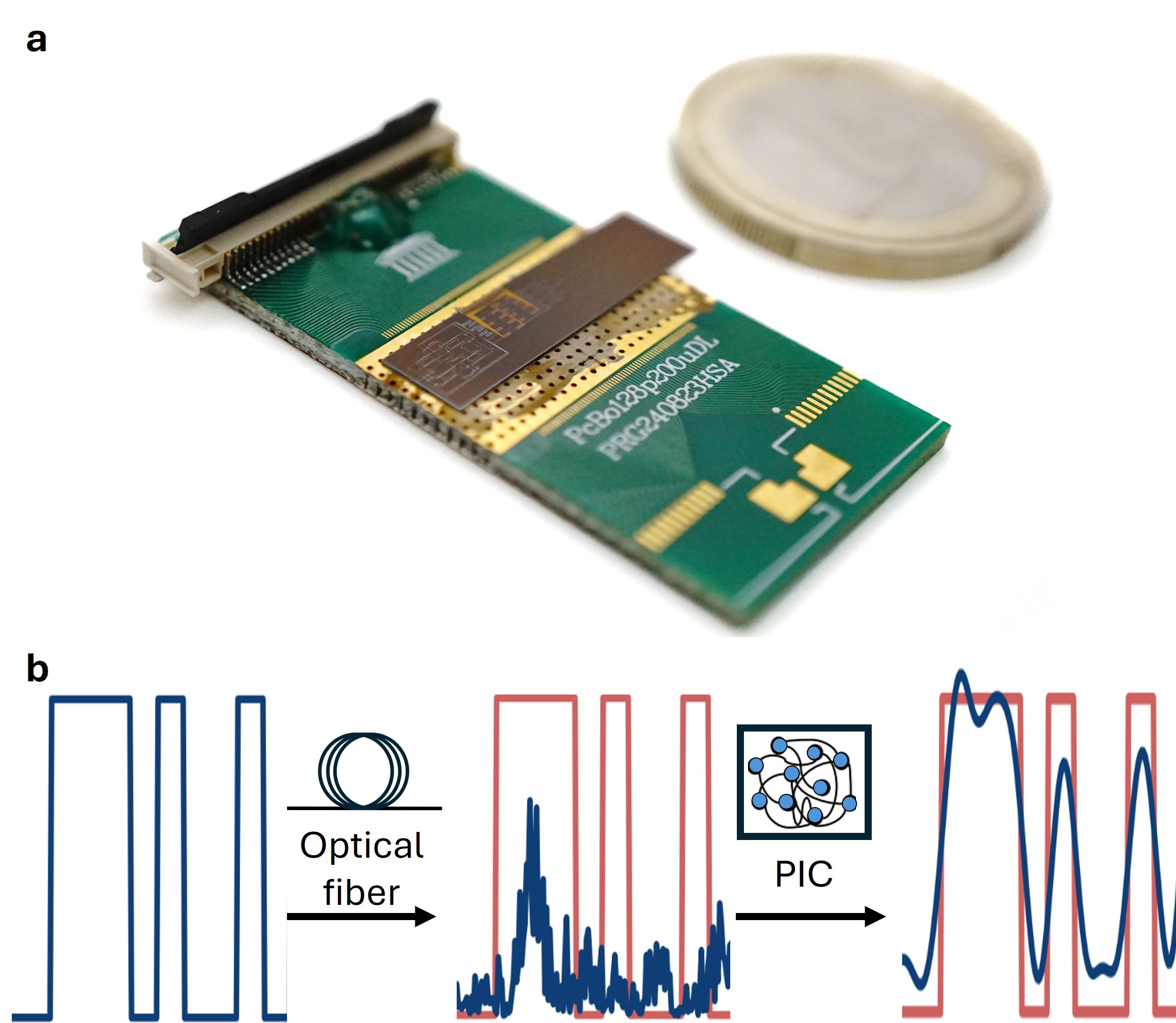}
        \caption{\textbf{Silicon photonics integrated reservoir} (a) Photograph of the photonic chip on a printed circuit board with a one-euro coin next to it for scale. (b) Diagram of signal equalisation using reservoir computer (PIC: Photonic Integrated Circuit).}
        \label{fig:eyecatch}
\end{figure}
Global data traffic has increased exponentially over the past twenty years \cite{dataneed}. Although optical networks were initially primarily designed for long-distance, high-speed transmission, emergent novel applications such as the Internet of Things, cloud services, artificial intelligence, and edge computing have drastically increased the amount of short-length inter- and intra-data center traffic. For long-reach optical networks, wavelength division multiplexing and coherent modulation formats are used frequently and can achieve high data rates and excellent spectral efficiency. Examples of these formats are  ``polarization division multiplexing quadrature phase shift keying'' (PDM-QPSK) and ``polarization division multiplexing quadrature amplitude modulation'' (PDM-QAM). However, the transceivers for these formats are costly, which makes them unsuitable for short-haul networks, where the rapid increase in demand for network capacity has made deployment and operational costs a critical factor \cite{dataneed2}. Currently,  ``intensity modulation with direct detection" (IM/DD) is the transceiver technology used for short-reach networks. However, this limits the achievable bit rate, since IM/DD only encodes data on the intensity of the signal, compared to coherent communication where information is encoded on the amplitude, phase, and/or polarization of the signal. To meet the increasing demands of the industry, it is essential to develop solutions that enhance data rates and power efficiency of IM/DD systems, while avoiding more costly coherent modulation formats.\\
When considering fibre-optic communication systems, a large constraint for the throughput of IM/DD systems stems from the distortions arising from propagation through optical fibres. These distortions can be subdivided into linear distortions, due to chromatic dispersion and polarization-mode dispersion, and nonlinear distortions due to the Kerr effect. These Kerr nonlinearities in particular pose challenges because of the self-phase modulation they induce. In a single channel, it is the interplay between chromatic dispersion and self-phase modulation that severely degrades the quality of the signal coming out of the fibre. State-of-the-art approaches utilize digital signal processing (DSP) techniques to mitigate these issues. Although effective, these algorithms introduce additional communication latency, as they require processing in the electrical domain. Moreover, their computational complexity leads to significant power consumption, presenting challenges for energy efficiency and operational costs in short-haul optical networks. Their performance for signals with high nonlinear distortions is limited as well \cite{DSP1,DSP2}. \\ 
Extensive research has been conducted to explore alternative methods for addressing these challenges. For compensating linear dispersion, solutions have been developed both in the electric and optical domains, many of which are now widely adopted. Examples include dispersion compensated/shifted fibres \cite{dispfib} and fibre Bragg gratings \cite{Bragg} in the optical domain, and fractionally spaced equalizers \cite{frac} and feed-forward equalizers \cite{ffe} in the digital electronic domain.\\
For the nonlinear part of the problem, electrical and optical solutions have been found as well. In the electric domain, digital backpropagation \cite{DBP} or nonlinear Volterra series \cite{Volt} are notable solutions. These implementations are power hungry however, due to the complexity of their algorithms, and introduce a latency due to their digital implementation. Optically the most important examples are optical phase conjugation (OPC) \cite{OPC} and phase-conjugated twin waves (PC-TW) \cite{TW}. However, OPC relies on an a priori known fibre length, which means it cannot be used for an all-purpose solution, and PC-TW sacrifices half the transmission capacity.\\
In addition to these implementations, research has been conducted to try and solve these problems using machine learning and neural networks, both in the digital electronic domain, as well as in the optical domain \cite{ML7, ML1,ML2,ML3,ML4,ML5,ML6}. An intuitive approach for machine learning in this context is to implement a neural network after the optical fibre or after the receiver and to consider this neural network as a post-processor that retrieves the original undistorted signal. One promising candidate for an optical implementation of this concept is reservoir computing (RC). This is a machine learning paradigm that uses a fixed, randomly initialized recurrent neural network, followed by a simple trainable linear readout. In practice, the random initialization typically stems from  fabrication imperfections. Moreover, the simple trainable linear readout has the advantage of being easy to train, as well as keeping power consumption limited.  \\ 
Within the reservoir paradigm, there are two possible implementations: time-delay-based reservoirs (TDRC) and spatially distributed reservoirs \cite{RCreview}. Experiments and simulations utilizing TDRCs have already shown the ability for optically equalizing dispersion \cite{RC1, RC2, RC3, RC4, RC5}. However, these systems require time multiplexing, which introduces challenges at high speeds due to the response time of the nonlinear neuron. This means that for TDRCs a trade-off exists between the computational capacity and the processing rate \cite{Ar1}\cite{Ar2}\cite{ESTE}. 
This issue is not present for spatially distributed reservoirs, where we showed previously that a linear and passive integrated circuit as a reservoir, in combination with a separate digital readout layer, forms a suitable solution for signal equalisation \cite{PaperStijn}. Additionally, in terms of bandwidth, a single chip with a single set of weights can simultaneously equalize signals at multiple wavelengths \cite{PaperEmmanuel}. However, both of these demonstrations did not have functional optical weights and all output channels needed to be measured sequentially and processed digitally offline. As a result, these systems did not operate in real time, despite the high-speed optical computation in the reservoir itself.  There was an initial attempt to integrate a readout layer, which however was limited in scale and only tested on simple academic benchmarks as opposed to on a complex, industrially relevant highly nonlinear task \cite{IntRead}.

In this work, we present a novel silicon photonics chip that incorporates a fully integrated optical readout layer, enabling real-time dispersion within a single-chip architecture with optical input and output. Our experiments demonstrate the chip's capability to equalize a 28 Gbps OOK signal at 10 dBm for fibre lengths up to 50 km and a 17 dBm signal over 25 km, achieving in all cases a bit error rate (BER) below the generic standard forward error correction (GFEC) threshold of $5.8 \times 10^{-5}$ \cite{FEC}.

This paper highlights several benefits of our integrated silicon photonics demonstrator:

\begin{enumerate}

    \item Real-Time Optical Signal Processing: 
    The chip processes signals online (i.e. in real time) entirely within the optical domain, eliminating the latencies associated with electronic computation or inter-domain connectivity as is the case for electro-optic solutions. The only latency is the intrinsic light propagation time within the chip, which is negligible in comparison. Importantly, the absence of electrical overhead further reduces power loss compared to electro-optic alternatives.

    \item High Modulation Speeds: The on-chip delay lines synchronize the reservoir’s time scale with the bit period. In the current design, we targetted compatibility with 28 GBd signals. However, as we will show, removing these delays allows compatibility with single-channel modulation rates nearing 900 GBd. Even though this is far beyond the capabilities of current transceivers and detectors, this illustrates that our approach is highly future-proof.

    \item Linear and Passive Device Design: 
    The reservoir comprises exclusively linear and passive components, such as multimode interferometers and waveguides. This reduces the fabrication and operational complexity, as integrated nonlinear devices are either difficult to integrate (e.g. if they require novel materials) or difficult to control (e.g. if they rely on resonances). The only required nonlinearity in our implementation is that of the final photodiode, which converts complex-valued fields to real-valued intensities, which is an intrinsically nonlinear operation \cite{Vandoorne}. Moreover, this nonlinearity does not require a minimum power level, paving the way to low power consumption. (Our current implementation uses integrated tungsten heaters, which act as complex weights. However, replacing these with nonvolatile weights, e.g. based on BTO \cite{BTOvol}, opens pathways to ultra-low power consumption.)
    
    \item Agnostic to Fibre Length and Input Power: 
    The chip demonstrates robust performance across diverse test cases involving both linear and nonlinear distortions. This was tested by experiments with fibre lengths up to 50 km for a 10 dBm input power into the fibre, and input powers up to 17 dBm for a 25 km fibre. This adaptability underscores its potential as a versatile solution for modern high-speed optical communication systems.

    \item CMOS Compatibility and Scalability:
    The chip is based on silicon photonics, ensuring compatibility with CMOS fabrication processes. This feature also enables seamless integration with silicon electronics and supports scalable production for widespread deployment and low cost.

\end{enumerate}

The rest of this paper is structured as follows. In Section 2, we will give a brief introduction to the reservoir computing formalism, followed by the design of the chip. Section 3 will explain the experimental results. Subsequently, section 4 discusses these results and provides a future outlook. Finally, section 5 concludes the paper.

\section{Integrated Photonic Reservoir Computing}
Reservoir computing is a recurrent neural network (RNN) approach introduced independently by Jaeger \cite{Jeger} and Maass \cite{Maass}. A reservoir computer is an RNN made up of three layers: \begin{enumerate}
\item an input layer, where (analog) signals are introduced to the reservoir.
\item the reservoir itself, typically a non-linear dynamical system.
\item an output layer, where signals processed by the reservoir undergo linear combination.
\end{enumerate}
Unlike traditional recurrent neural networks, where all interconnections are trained, reservoir computing retains fixed internal connections inside the reservoir. The random and static nature of the reservoir network in RC makes it well-suited for analog computing, as it relaxes fabrication tolerances and complexity compared to physical RNNs.\\
\begin{figure*}[h!]
     \centering
     
         \centering
         \includegraphics[width=\textwidth]{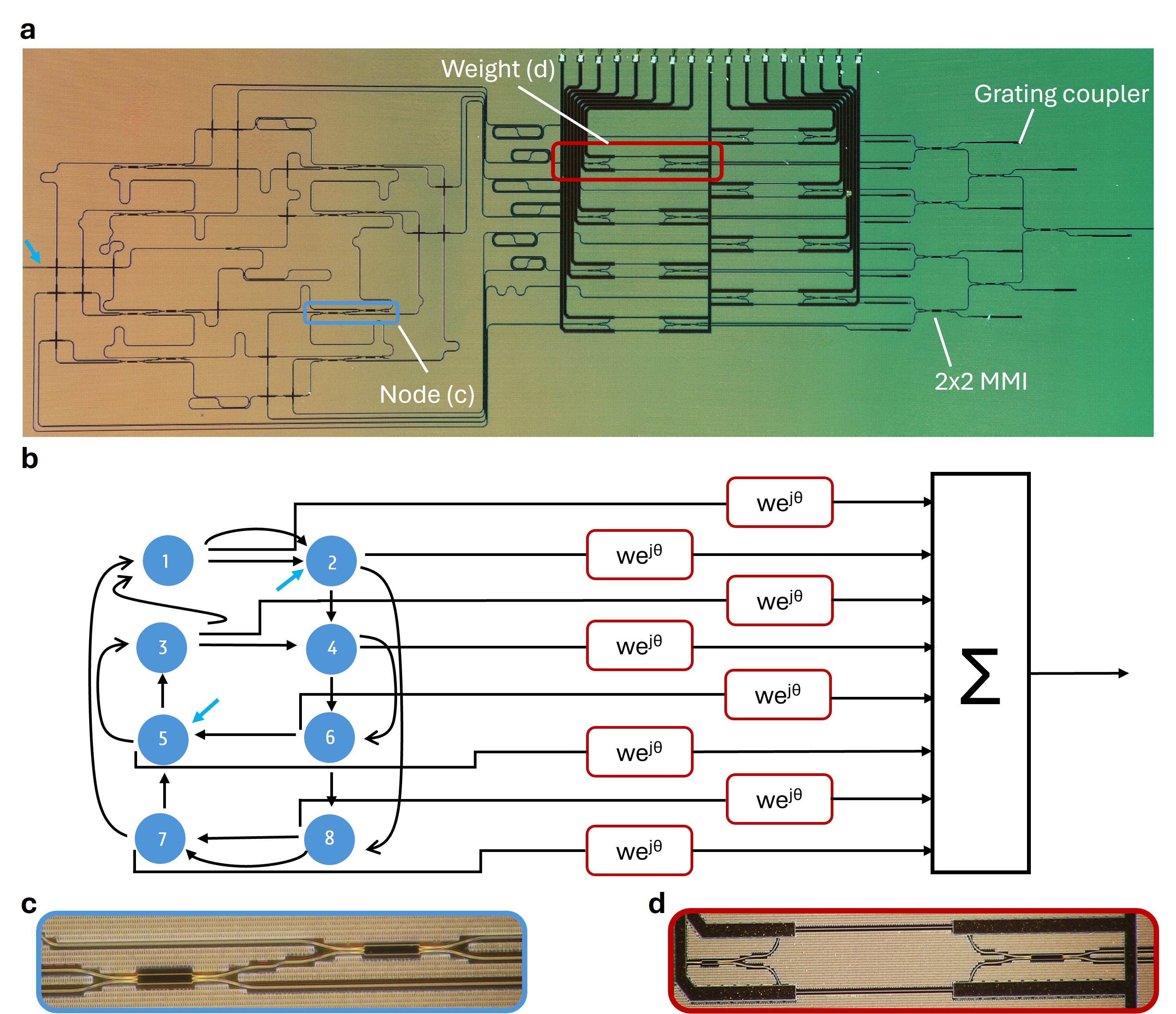}

        \caption{\textbf{Design of photonic integrated reservoir} (a) Microscope image of the chip design with a blue rectangle showing a node, and a red rectangle showing a weight implemented as an MZI with heaters. (b) Schematic representation of the chip. Blue circles indicate nodes, and arrows indicate the propagation direction of the light. Note that the detector is off-chip (c) Close-up image of a node composed of two 2x2 MMIs. (d) Close-up image of an MZI with heaters.}
        \label{fig:Chips}
\end{figure*}

Our photonic RC architecture is implemented in a Silicon Nitride platform and follows a four-port design \cite{4port}. At each output of the reservoir, the signal will be a superposition of many different signals with different delays and different phases. In practice, the randomness of the reservoir is caused by fabrication-induced  variations in the surface roughness and in the length of the waveguides. This results in a random (but fixed for any given chip) set of phase changes for the propagating signals. The nonlinearity of the photodiode, which acts as the receiver, will transform the relative phases of the different interfering signals into power variations. It turns out that this nonlinearity by itself gives us enough processing power to perform signal equalisation, as shown in previous experiments \cite{PaperStijn}\cite{PaperEmmanuel}.
The eight nodes of the reservoir are implemented using multi-mode interferometers (MMIs), with passive waveguides forming the interconnections. A microscope image of the fabricated chip is shown in Figure \ref{fig:Chips}a, while a schematic representation is provided in Figure \ref{fig:Chips}b. The four-port architecture usually uses 3x3 MMIs as nodes. Indeed, each node features three inputs and three outputs: one input connects to the input layer, one output links to the readout layer, and the remaining ports facilitate interconnections between nodes. However, in order to be able to use more standard building blocks, in our current design each node is constructed by cascading two 2×2 MMIs (Figure \ref{fig:Chips}c), rather than using a single 3x3 MMI. This approach introduces asymmetry in the node transfer functions, but this can enhance RNN performance.  The input layer distributes the chip’s single input to two reservoir nodes. Two inputs for the reservoir are used instead of one, to have a more even signal distribution across the network.\\
A key design parameter is the length of the interconnecting waveguides, or delay lines, which must be tuned to the data rate of the signal. Simulations indicate that an interconnection delay of half a bit period optimally balances signal propagation.\\
The readout layer consists of eight Mach-Zehnder interferometers (MZIs), each corresponding to a reservoir node and tuned via two thermoelectric heaters (Figure \ref{fig:Chips}d). Adjusting the heater currents modifies the local temperature, thereby controlling both  amplitude and phase of each MZI’s output. This enables weighting of the reservoir signals before they are recombined in a summation tree. 
Due to design and fabrication imperfections, the chip introduces significant losses, measured at 30 dB end-to-end in our implementation. Integrating a semiconductor optical amplifier (SOA) could mitigate this loss while introducing additional nonlinearities, which may enhance performance for complex tasks \cite{SOA}. 

\begin{figure*}[h!]
     \centering
     
         \includegraphics[width=\textwidth]{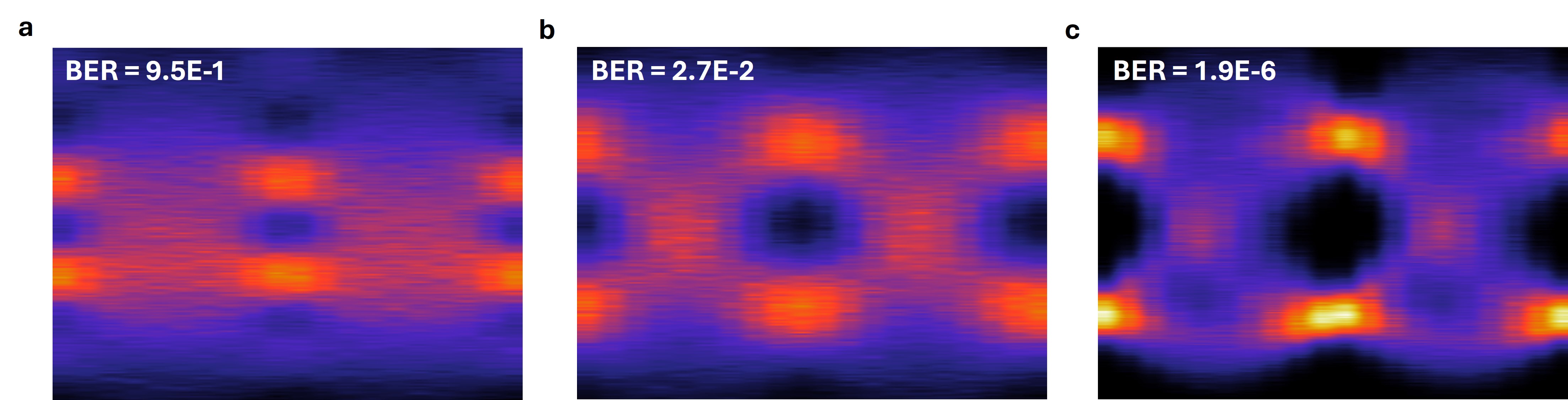}
        \caption{\textbf{Experimental eye diagrams} Eye diagrams and corresponding statistical bit error rates of (a) the distorted signal after propagating through 50 km of optical fibre with an input power of 10 dBm, (b) after applying a digital feed-forward equalizer with 17 taps, and (c) after applying the reservoir chip.}
        \label{fig:eye}
\end{figure*}

\begin{figure*}
     \centering
     
         \centering
         \includegraphics[width=\textwidth]{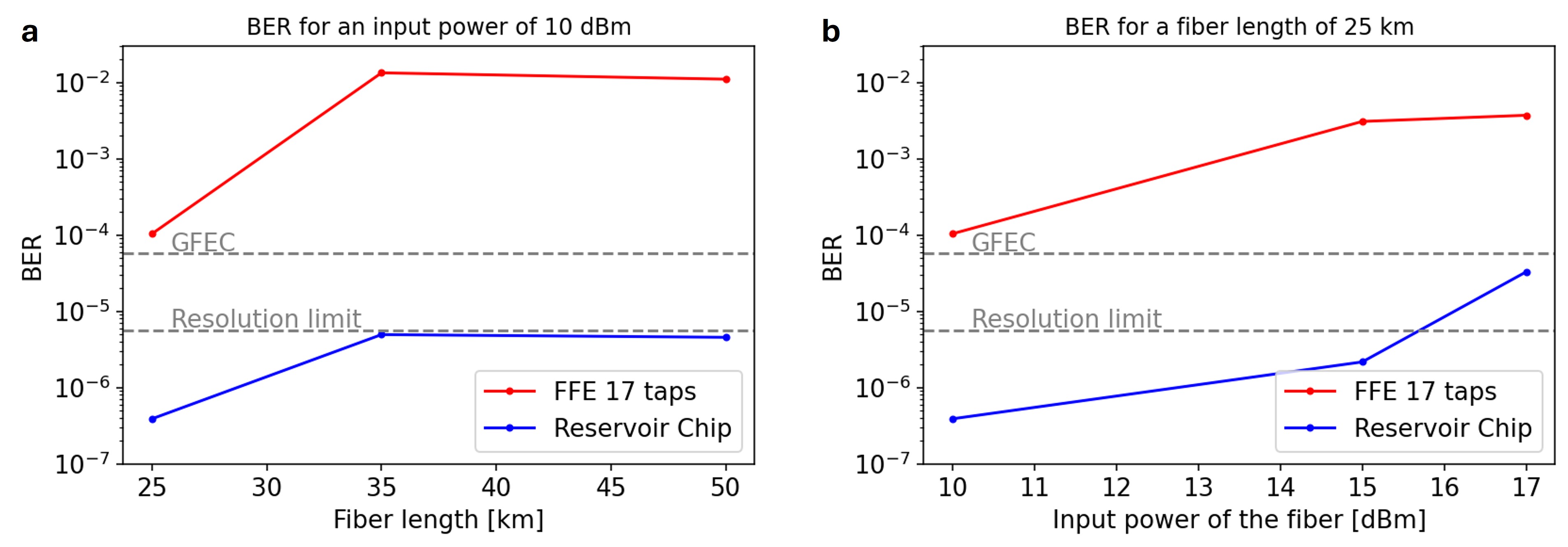}

        \caption{\textbf{Experimental bit error rates} Measured bit error rates  for (a) different lengths of the optical fibre for an input power of 10 dBm and (b) different input powers for a fibre length of 25 km. The grey dashed line signifies the resolution limit up to which predictions can accurately be made for the reservoir, and the GFEC (Generic forward error correction) boundary.}
        \label{fig:res}
\end{figure*}

\section{Experimental results}
The experiments evaluated the system's performance while varying two parameters: the input power to the fibre and the fibre length. The input power was swept from 10 to 17 dBm for a fixed fibre length of 25 km, while the fibre length was varied from 25 to 50 km at a constant input power of 10 dBm. An increase in fibre length extends the interaction distance for both linear and nonlinear dispersion effects, whereas increasing the input power primarily enhances the impact of the Kerr nonlinearity, thereby increasing the nonlinear complexity of the machine learning task. For the complete set-up and processing steps, we refer to the Methods section.

Figure \ref{fig:eye} provides a visual comparison of the equalisation effectiveness through eye diagrams. These diagrams illustrate the signal degradation and subsequent restoration for a transmission distance of 50 km at an input power of 10 dBm.  Each eye diagram corresponds to 90,000 bits.
The first eye diagram, corresponding to the raw distorted signal at the output of the fibre, exhibits significant inter-symbol interference (ISI) due to chromatic dispersion, leading to severe eye closure. The second diagram shows a partially restored bitstream after applying a digital feedforward equaliser (FFE). However, residual nonlinear distortion effects remain, and the overall eye opening is still limited. In contrast, the third diagram - after equalisation by the photonic reservoir — demonstrates a significantly wider eye opening, because in contrast to the FFE, the reservoir can also perform nonlinear operations. Not only is the black region of the eye wider than for the FFE, signifying a well-defined transition region, the eye is higher as well, indicating that the overall distinction between logical ones and zeros is considerably improved. 

The eye diagrams also show the statistical BER, as calculated based on the assumption of a Gaussian distribution of ones and zeros. This serves as an initial quality metric for the eye diagrams. However, for subsequent plots, we will show the more accurate BER obtained through direct error counting over a larger dataset.

These BER results for different experimental conditions are summarized in Figure \ref{fig:res}. Figure \ref{fig:res}(a) presents BER trends as fibre length increases at a fixed input power of 10 dBm. In this regime, chromatic dispersion is the dominant performance-limiting factor. For longer fibre spans, performance degrades due to the limited memory of both equalisation systems. Notably, despite the FFE possessing a larger memory capacity than the photonic reservoir, the reservoir outperforms the FFE by at least two orders of magnitude in BER at any fibre length. This is because the reservoir performs rich operations on the input, compared to the FFE, where only linear combinations of delayed versions of the same signal are used. Another advantage compared to the FFE is that the reservoir processes the signal before the photodetector, avoiding problems due to frequency fading that digital implementations suffer from.

Figure \ref{fig:res}(b) shows the BER evolution as input power increases for a fixed fibre length of 25 km. In this case, nonlinear dispersion and its interaction with chromatic dispersion impose performance constraints. Even with the same number of degrees of freedom, the photonic reservoir again surpasses the FFE, achieving at least a two-order-of-magnitude improvement in BER.

In both experimental conditions, the photonic reservoir achieves equalisation well below the generic forward error correction (GFEC) limit \cite{FEC} across all tested configurations, whereas the FFE fails to do so. This implies that the reservoir-based system requires a significantly less computationally intensive forward error correction (FEC) stage compared to the FFE. For a more general comparison with state-of-the-art optical implementations of equalisation, we refer to Table~\ref{tab:comparison}.

\begin{table*}[h!]
    \footnotesize
    \centering
    \caption{Comparison of different state-of-the-art experimental optical equalisation solutions for intensity modulation \cite{Review1}\cite{review2}. Fibre length and launch power are included, since higher values for these parameters result in a more nonlinear task that is drastically more difficult to solve. Most experiments are not real-time, in the sense that either the input signal needed to be buffered and slowed down, or that the output signals were captured and subsequently processed off-line on a PC.\\
    (B: Datarate, L: Fibre Length, P: Launch Power)}
    \begin{tabular}{|c|c|c|c|c|c|c|}
        \hline
        \textbf{Reference} & \textbf{B [GBaud]} & \textbf{Modulation Format} & \textbf{L [km]} & \textbf{P [dBm]} & \textbf{Real Time} & \textbf{BER*} \\
        \hline
        A. Argyris \cite{Ar1} & 25 & PAM-2 & 45  & 10 & No  & \num{1.8e-4} \\
        \hline
        A. Argyris \cite{Ar2} & 28 & PAM-4 & 27  & 14 & No  & \num{2.0e-3} \\
        \hline
        S. Sackesyn \cite{PaperStijn}   & 32 & PAM-2 & 25  & 18 & No  & \num{1.0e-3} \\
        \hline
        I. Estébanez \cite{ESTE}  & 56 & PAM-4 & 100 & 6  & No  & \num{3.0e-4} \\
        \hline
        S. Li \cite{Ring}\textsuperscript{1} & 56 & PAM-4 & 100 & 0 & No & \num{<3.0e-4}\\
        \hline
        G. Sarantoglou \cite{FP} & 28 & PAM-4 & 40 & - & No & \num{4.2e-2}\\
        \hline
        K. Sozos \cite{RossT}  & 64 & PAM-4 & 50 & 3  & Yes  & \num{<1.0e-2} \\
        \hline
        E. Staffoli \cite{Percep}\textsuperscript{2}  & 10 & PAM-2 & 125 & 0  & Yes & $<$\num{1.0e-4} \\
        \hline
        This work & 28 & PAM-2 & 25 & 17 & Yes & $<$\num{5.6e-6} \\
        \hline
        This work & 28 & PAM-2 & 50 & 10 & Yes & \num{5.7e-5} \\
        \hline
    \end{tabular}
    
    \label{tab:comparison}
\end{table*}

While all but one BER measurement of the reservoir falls below the resolution limit of the experimental setup, further increasing the test set size was not pursued due to hardware constraints. Indeed, the arbitrary waveform generator (AWG) used in these experiments has a limited bitstream capacity per transmission cycle of approximately $10^5$ bits. Achieving a resolution limit of $10^7$ would require a test set of $10^9$ bits, necessitating $10^4$ AWG transmission cycles, compared to the 200 cycles performed for each experiment in this study. However, the close agreement between the statistical BER estimation and the directly measured BER strongly suggests that additional measurements would yield minimal deviation from the reported results.

\section{Future Outlook}

The presented platform demonstrates significant potential for future advancements, particularly in terms of throughput and power consumption. Simulations based on the current work (see Supplementary Materials) indicate that simply by eliminating the on-chip delay lines, the reservoir becomes compatible with a data rate of 896 Gbps. Although no modulators or detectors support these data rates, simulations show that hypothetically this platform could perform dispersion compensation over a distance of 24 m at a launch power of 10 dBm. (This configuration has a  $B^2 \times L$ product approximately equivalent to our experiments of a 28 Gbps signal transmitted over 25 km.) These results underline the platform's capacity to support future innovations in single-channel transceiver technology, by trivial modifications of the architecture.\footnotetext{\textsuperscript{1}A number was not provided in the paper, therefore an estimation of the best result was made based on the provided graphs.}
\footnotetext{\textsuperscript{2}The provided value is adjusted for the size of the dataset according to the resolution limit of 100 bits.} 

To estimate the ultimate throughput of this design, we consider the combination of these simulations with prior experimental and simulation results related to an electric readout implementation of the same architecture for wavelength-division multiplexing \cite{PaperEmmanuel, Gooskens}. These studies have demonstrated that the four-port architecture can use the same set of weights to separately equalise tens to hundreds of distinct wavelength channels. Assuming the utilization of 100 channels, the achievable aggregate throughput would amount to 89.6 Tbps. In terms of metrics like TOPS (tera-operations per second), the number would be even higher, since inside the network each bit undergoes many elementary operations like splitting, combining and weighting. However, it is neither straightforward nor advisable to use this metric in this case, given the analog nature of the computations and the recurrent character of the architecture.

Regarding power consumption, two critical aspects of the current setup must be addressed: pre- and post-amplification on-chip and the use of power-efficient weighting mechanisms. 

Pre-amplification can be efficiently realized using a semiconductor optical amplifier (SOA) at the chip's input, requiring approximately 540 mW to achieve a signal gain of 15 dBm from an input power of 0~dBm \cite{SOA}. Post-amplification may be rendered unnecessary by employing sensitive integrated detectors in future designs. Notably, the power budget of these photodiodes is excluded, as photodiodes would be present irrespective of the inclusion of the equalizing photonic reservoir. Finally, note that there is still considerable room to improve the insertion losses of the chip itself, as the current design still relied in part on grating couplers rather than edge-couplers, and as the MMIs were slightly off-spec.

For weighting, non-volatile phase modulators, such as barium titanate (BTO), represent a promising approach. These modulators, once trained, exhibit minimal static power consumption, since the leakage current has been demonstrated to be as low as 100 nW for devices operating at 25 GHz \cite{LeakPow}. It is noteworthy that the speed of these modulators need not match the data rate, as they remain static after training. Consequently, the total potential power consumption of the chip is estimated to be approximately 540 mW, i.e. entirely dominated by the SOA.

Based on these metrics, the platform's power consumption per bit can be estimated through dividing the projected power consumption by the throughput.  Using the two different throughput values provided earlier, the power consumption is projected to be  558 fJ/bit for a single channel or 5.58 fJ/bit for the wavelength-multiplexed system, without requiring electro-optic interfacing during inference. 

These findings highlight the potential of this architecture for low-power equalisation at ultrahigh data rates. Furthermore, the architecture's inherent versatility supports various modulation formats, including those used in coherent and self-coherent communication systems, as we have previously shown in  \cite{PaperSarah}.  This demonstrates its adaptability across different data rates, wavelengths, and modulation schemes.

\section{Conclusion}
The cost-efficient and accurate mitigation of linear and nonlinear dispersion in fibre-optic communication remains a longstanding challenge in both industry and research. In this work, we experimentally demonstrate a photonic reservoir computing chip capable of equalizing a 28 Gbps on-off keying (OOK) signal well below the generic forward error correction (GFEC) threshold of $5.8\times10^5$ for fibre lengths up to 50 km and input powers up to 17 dBm. Furthermore, we establish the superiority of this approach over a conventional digital feed-forward equalisation (FFE) method, which performs at least two orders of magnitude worse, and it particularly fails in reaching the same GFEC limit. Crucially, this architecture enables signal equalisation across different modulation formats and wavelengths while preserving an optical output, distinguishing it from many electro-optic alternatives that require optical-to-electrical conversion. Simply by eliminating on-chip delay lines and moving to non-volatile phase shifters, our projections show a possible throughput of 89.6 Tbps at a  power consumption of 5.58 fJ/bit.

\section{Acknowledgements}
This work was supported by the European Commission in the Horizon Europe programme under the projects Nebula, Prometheus, Neho and NEHIL.

\printbibliography
\newpage
\section{Methods}
\subsection{Setup}
\begin{figure*}[h!]
     \centering
     
         \centering
        \includegraphics[width=1\textwidth]{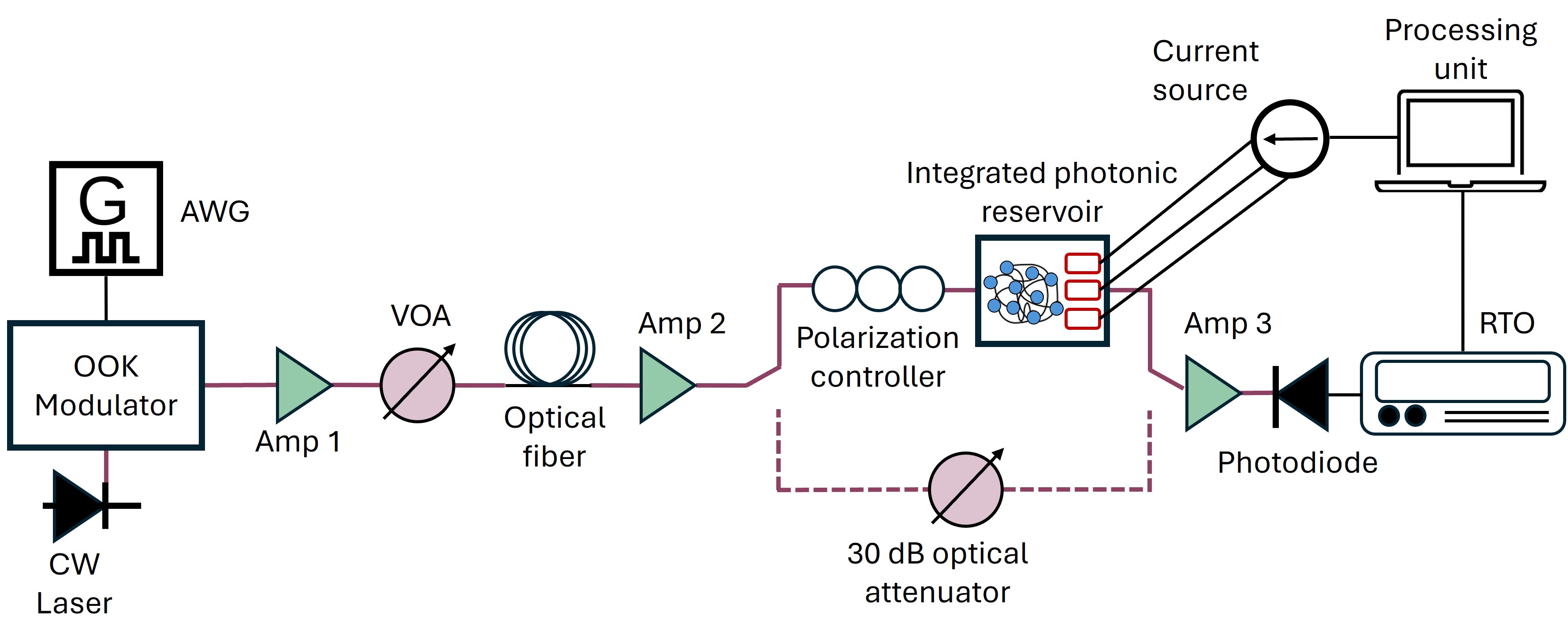}
        \caption{\textbf{Experimental set-up} The two different setups for the experiments. The upper and lower path respectively show the setup with and without the photonic reservoir. (CW: Continuous wave, AWG: Arbitrary waveform generator, OOK: On-Off keyeing, Amp: Amplifier, RTO: real-time-oscilloscope)}
        \label{fig:setup}
\end{figure*}
The experimental setup is depicted in Figure \ref{fig:setup}. A 1550 nm continuous wave (CW) laser is modulated using an on-off keying (OOK) modulator to generate a 28 Gbps intensity-modulated optical bitstream. To generate this bitstream, the Mersenne Twister algorithm is used as a generator, as opposed to a PRBS signal. This is done to avoid that the systems learns the PRBS rule being used instead of being able to equalise the actual signal. As such training and testing data is generated using different initialisation seeds. To precisely control the input power without affecting the system's noise figure, the modulated signal first undergoes a fixed  amplification to 20 dBm, followed by a tuneable attenuation. The amplification is provided by an erbium-doped fibre amplifier (EDFA).
As the signal propagates through the optical fibre, dispersion introduces distortions that necessitate compensation. Because of design and fabrication imperfections, the photonic integrated circuit experiences an insertion loss of 30 dB, and therefore the signal needs to be amplified accordingly. Performance is optimal when this is done through both pre-and post amplification. EDFA 2 will amplify the signal up to 15 dBm, while EDFA 3 will amplify the output of the reservoir to 10 dBm, an optimal power level for the high-speed photodetector. The photocurrent of this detector is captured by a real-time oscilloscope operating at 160 GSamples/s. The acquired data is processed using an iterative algorithm (to be discussed later) that dynamically adjusts the driving currents for the PIC’s weight elements, forming a closed-loop optimization cycle.
A critical component of the setup is polarization control. The photonic circuit integrates both edge couplers and grating couplers, the latter being highly polarization-sensitive. Additionally, on-chip waveguides exhibit polarization-dependent propagation characteristics, with transverse magnetic (TM) modes suffering from significantly higher losses. To mitigate polarization-induced performance degradation, a polarization controller is employed prior to the circuit, ensuring that the input signal is appropriately aligned for optimal transmission through the device. A list of all model numbers is given in table \ref{tab:equip}.

\begin{table*}[h!]
    \centering
    \caption{Equipment used in the set-up and their corresponding model number}
    \begin{tabular}{|l|l|}
        \hline
        \textbf{Equipment} & \textbf{Model} \\ \hline
        AWG & Keysight M8195a \\ \hline
        RTO & Teledyne LeCroy SDA 830Zi-B \\ \hline
        Laser & CoBrite DX1 S type \\ \hline
        Modulator & Exail MX-LN-40 \\ \hline
        Amplifier 1 & Keopsys CEFA-C-HG-SM-50-B201-FA-FA \\ \hline
        Amplifier 2 \& 3 & Keopsys CEFA-C-HG-SM-50-B130-FA-FA \\ \hline
    \end{tabular}
    \label{tab:equip}
    
\end{table*}

To enable a direct performance comparison between the photonic system and an electrical feed-forward equalizer (FFE) based on a tapped delay line, it is essential to ensure that the signals in both scenarios exhibit equivalent signal-to-noise ratios (SNR). To achieve this, a secondary signal path is introduced, as illustrated in Figure \ref{fig:setup}, where the photonic reservoir is replaced by a 30 dB optical attenuator. This configuration maintains comparable power levels while eliminating the impact of the photonic reservoir's transformation. Furthermore, all post-processing techniques applied to the reservoir output are identically applied to this secondary benchmark signal path, ensuring an unbiased evaluation of system performance.
\subsection{Post-Processing}
The data processing pipeline is different for the two succesive key stages in the experiment: (i) training, which determines the optimal currents for driving the weight elements on the photonic chip, and subsequently (ii) inference, where the received signal is processed by the reservoir using the weights deterimined in the training step. It is important to emphasize that all processing steps employed in this work beyond the reservoir are standard techniques in intensity-modulation direct-detection (IM/DD) systems and are not introduced specifically due to the presence of the photonic chip.

\subsubsection{Training}
\begin{figure*}[h!]
     \centering
     
         \centering
         \includegraphics[width=1\textwidth]{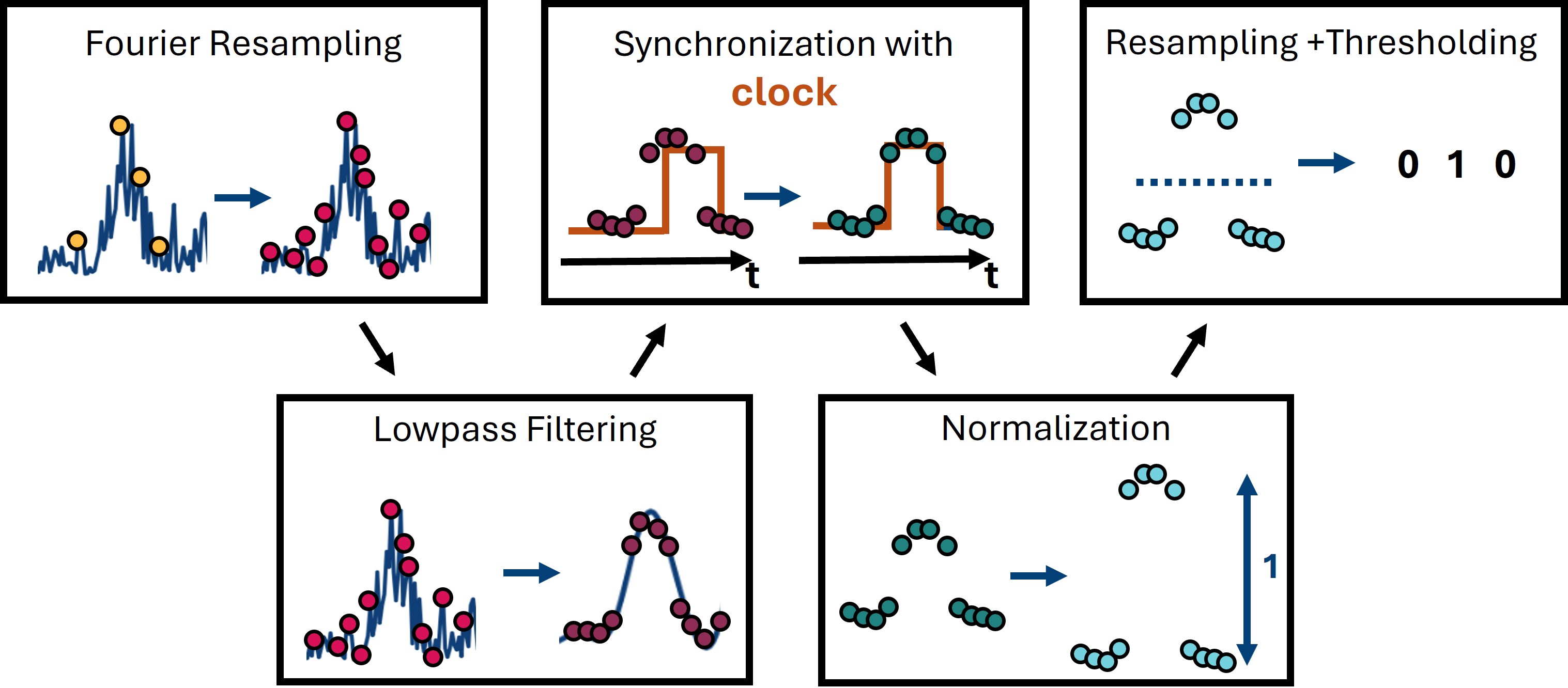}
        \caption{\textbf{Post-processing methods} Schematic showing the processing steps taken after receiving the signal. Note that filtering and normalization only take place during inference and are not used during training.}
        \label{fig:proces}
\end{figure*}

During training, we send a known pattern of 20,000 bits through the system and adapt the weights of the system in such a way that the error rate is minimal. A conceptual schematic of the data processing steps is presented in Figure \ref{fig:proces}. Fourier resampling, synchronization and thresholding are performed both during training and inference. Filtering and normalization take place only during inference. Due to the non-integer relationship between the real-time oscilloscope (RTO) sampling rate (160 GSamples/s) and the data rate (28 Gbps), the signal is first resampled via numpy's Fourier interpolation to ensure a consistent sampling grid aligned with the bit period. Note that in an actual system with an integrated receiver, this step would not be necessary. The resampled signal is then synchronized with the known training pattern we sent through the link using a predefined header consisting of 1000 consecutive zero bits at the start of the transmission. This is important, since we are constantly sending data to the oscilloscope, but without this synchronisation, the machine learning algorithm would not be able to tell what the first bit of the signal is, and therefore it would not have the correct labels to perform the training. Once synchronization is achieved, the signal is thresholded to convert it into a bitstring. Finally, the bit error rate (BER) is computed through straightforward counting as the primary performance metric.

To optimize the readout layer weights, we employ the covariance matrix adaptation evolution strategy (CMA-ES) \cite{cmaes}, a robust evolutionary optimization algorithm designed to efficiently refine high-dimensional parameter spaces. The optimization process spans only 750 iterations, structured into 15 generations with a population size of 50. To accelerate convergence, the standard deviation of the algorithm is manually tuned from an initial value of 3 to a final value of 0.3, progressively narrowing the search space. We verified that the 20,000 bits that are used for training in these experiments are sufficient to ensure a well-conditioned model for subsequent inference.

\subsubsection{Inference}
During inference, additional processing steps are performed to ensure an accurate estimation of the system's bit error rate (BER). The signal first undergoes resampling using the same methodology as in the training phase, followed by low-pass filtering with a bandwidth equal to the data rate (28 GHz) to suppress high-frequency noise. After filtering, synchronization is carried out analogously to the training stage, in order to be able to determine the ground truth for the received bits. The signal is then normalized by subtracting the mean and dividing by the standard deviation, a process performed in packets of 1,000 received bits to account for system fluctuations.
The final processing step is thresholding. Unlike in the training phase, where the training signal itself is used to determine the optimal decision boundary, inference relies on a separate validation dataset to establish a fixed threshold, which is then applied to all test data. This threshold is determined using a validation set of 1,000 bits. Once thresholding is completed, the BER is computed by comparing the predicted bitstream with the target sequence and counting the errors. Each experiment utilizes a test set comprising 18 million bits. Due to statistical variations, BER estimation is subject to a resolution limit, which according to rules-of-thumb \cite{BERest} corresponds to a minimum of 100 erroneous bits, or 5.55e-6 for these experiments.\\
As discussed in the main text, the photonic reservoir computing system is benchmarked against a digital FFE. The FFE is a linear filter that constructs a weighted sum of the signal and its delayed copies, with its memory depth determined by the number of taps. Often in filter design an odd number of taps is chosen, since for an FFE this means that the final delay is an integer number of samples, and all frequencies are delayed in the same way. Therefore, in this study, an FFE with 17 taps is employed, ensuring a fair comparison, as the photonic reservoir system utilizes only 16 trainable weights. It is also worth pointing out that even an FFE with 16 taps would still exhibit a greater memory capacity than the photonic reservoir because of optical losses, making sure that we do not give a preferential treatment to the reservoir in the comparison.
To further maximise the performance of the baseline, the latency between the FFE output and the target sequence is separately optimized. Note that this delay was not optimized for the reservoir, where we match the headers of the measured output signal and the target without adding extra delay. In order to find the optimal weights for the digital filter and threshold, a validation set is used. Given that the FFE relies on linear regression for weight optimization, a larger validation dataset is required to mitigate underfitting compared to the reservoir case where the validation set is only used for thresholding. Consequently, the validation set size for the FFE is increased to 4,000 bits, compared to 1,000 bits for the photonic reservoir system.
\clearpage
\section{Supplementary Materials}
\subsection{Scaling simulations}
\begin{figure*}
     \centering
     
         \centering
         \includegraphics[width=\textwidth]{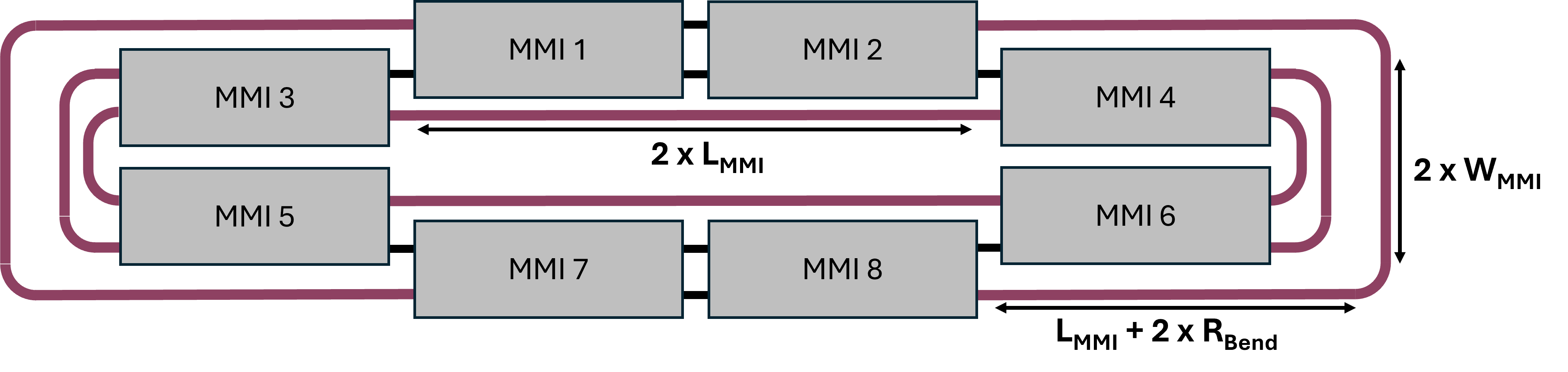}

        \caption{\textbf{Architecture for high-speed simulations} Schematic representation of the four-port architecture for high data rates. The waveguides in black can have an arbitrary length, the waveguides in red are restricted by the dimensions of the MMI.}
        \label{fig:schem}
\end{figure*}
\begin{figure}
     \centering
     
         \centering
         \includegraphics[width=0.5\textwidth]{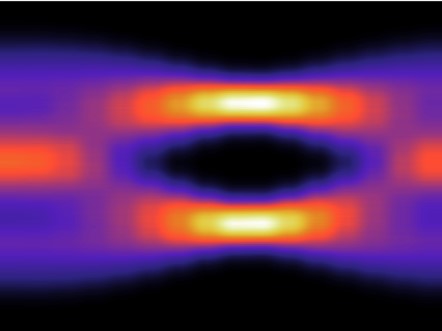}

        \caption{\textbf{Simulated eye diagram of equalized 896 GHz signal}}
        \label{fig:eyehs}
\end{figure}

When scaling the waveguides of the four-port reservoir to higher data rates, the waveguide lengths must be reduced to maintain an equivalent time delay for each bit within the reservoir system. However, the dimensions of the multimode interference couplers (MMIs) impose fundamental limits on the minimum achievable waveguide length. An optimal configuration for the four-port architecture, allowing the largest number of waveguides to be arbitrarily shortened, is depicted in Figure \ref{fig:schem}. In this figure, black lines represent waveguides that can be made arbitrarily short, while red lines indicate waveguides with a fixed minimum length. The most favorable MMI configuration utilizes 3x3 MMIs rather than cascaded 2x2 MMIs, with assumed dimensions of 540 µm in length and 50 µm in width. Additionally, bend radii are assumed to be 50 µm.\\
A critical question arises regarding whether such an architecture, accounting for these physical constraints, can at these higher bit rates still achieve dispersion compensation comparable to that demonstrated experimentally. To evaluate this, we investigate a bit rate of 896 Gbps over a transmission distance of 24 m. This configuration has a  $B^2 \times L$ product approximately equivalent to a 28 Gbps signal transmitted over 25 km, as investigated experimentally in this work. Note that in contrast to the more common $B \times L$ product, we are interested in situations with equal dispersion. When doubling the datarate, we should divide the fiber length by a factor of four to attain a configuration with equal chromatic dispersion.
To isolate the reservoir's intrinsic capabilities, we assume an ideal performance of the devices in the perifery: the modulator is modeled as producing a bitstream without impairments, the readout layer weights are presumed to function perfectly without frequency-related issues and the detector is also assumed to have infinite bandwidth. Simulations are averaged over 10 different random waveguide phase initialisations to ensure robustness against fabrication variability.
The simulation of the modulated signal and dispersion effects is conducted using VPIphotonics Design Suite \cite{VPI}, while the photonic reservoir is modeled using the Python library Photontorch \cite{photo}, which employs an S-matrix-based simulation method. Training and testing are conducted on bitstreams of $2^{14}$ and $2^{21}$ bits, respectively, ensuring that the resolution limit of 100 bits remains below the generalized forward error correction (GFEC) limit.
A representative eye diagram from the simulations is shown in Figure~\ref{fig:eyehs}. Without the reservoir, the BER for the chosen parameters amounts to \num{26e-2}. After applying our reservoir however, the calculated bit error rate (BER) amounts to the resolution limit of \num{4.77e-5}, which is below the GFEC limit of $5.8 \times 10^{-5}$. This result demonstrates that the reservoir architecture can effectively equalize signals even at high bit rates, despite the imposed physical constraints.

\end{document}